\documentclass[12pt,preprint]{aastex}

\shorttitle{Gravitational Field in the Outer Solar System}
\shortauthors{Page, Dixon, \& Wallin}

\begin{document}

\title{Can Minor Planets be Used to Assess Gravity in the Outer Solar System?}

\author{Gary L. Page\altaffilmark{1}, David S. Dixon\altaffilmark{2}, \and John F. Wallin\altaffilmark{3}}


\altaffiltext{1}{George Mason University, School of Computational Sciences, Center for Earth Observing and
                 Space Research (CEOSR), 4400 University Drive, MS 5C3, Fairfax, VA 22030; gpage@gmu.edu.}
\altaffiltext{2}{Jornada Observatory, Las Cruces, NM; ddixon@cybermesa.com.}
\altaffiltext{3}{George Mason University, School of Computational Sciences, Center for Earth Observing and
                 Space Research (CEOSR), 4400 University Drive, MS 5C3, Fairfax, VA 22030; jwallin@gmu.edu.}

\begin{abstract}
The twin Pioneer spacecraft have been tracked for over thirty years as they headed out of the solar system. 
After passing 20 AU from the Sun, both exhibited a systematic error in their trajectories that can be interpreted 
as a constant acceleration towards the Sun. This Pioneer Effect is most likely explained by spacecraft systematics, 
but there have been no convincing arguments that that is the case. The alternative is that the Pioneer Effect 
represents a real phenomenon and perhaps new physics. What is lacking is a means of measuring the effect, its 
variation, its potential anisotropies, and its region of influence. We show that minor planets provide an 
observational vehicle for investigating the gravitational field in the outer solar system, and that a sustained 
observation campaign against properly chosen minor planets could confirm or refute the existence of the Pioneer 
Effect. Additionally, even if the Pioneer Effect does not represent a new physical phenomenon, minor planets can 
be used to probe the gravitational field in the outer Solar System and since there are very few intermediate range 
tests of gravity at the multiple AU distance scale, this is a worthwhile endeavor in its own right. 
\end{abstract}

\keywords{astrometry; celestial mechanics; ephemerides; interplanetary medium; minor planets, asteroids; 
          solar system: general}

\section{INTRODUCTION}
Beginning in 1980 when Pioneer 10 was 20 AU from the Sun, analysis of unmodeled accelerations found 
that the biggest systematic error in the acceleration residuals was a constant acceleration towards the 
Sun of approximately 8$\times$10$^{-8}$ cm sec$^{-2}$ \citep{and98}. When Pioneer 11 passed this 20 
AU threshold, a similar effect was seen. Prompted by this unusual result, Galileo and Ulysses data were 
investigated for a similar effect. Although the limited data available from Galileo could not be used, 
Ulysses showed a similar unmodeled acceleration residual, even at its much smaller heliocentric 
distance \citep{and98}. The effect on the Pioneers has persisted until at least a heliocentric distance 
of 75 AU. 

\citet{and02a} discusses  a large number of potential causes of the anomalous acceleration, ranging 
from measurement methodology errors and gas leaks to modeling deficiencies and electromagnetic forces. 
The paper reviews a number of attempts to explain the anomalous acceleration in terms of known physics, 
and continues by reviewing a large number of potential explanations for the anomalous acceleration in 
terms of new physics. These include: whether the effect is due to dark matter or a modification of 
gravity; whether it is a measure of spacetime curvature and cosmological expansion; and whether it is 
due to a number of more radical variants on the relativistic gravity theme. 

In the end, \citet{and02a} finds ``no mechanism or theory that explains the anomalous acceleration.'' 
Thus, in the minds of those authors, the possibility of new physics should not be ruled out. Interest 
in this phenomenon continues. For example, \citet{and02b} reports a potential consequence of a Pioneer 
effect in the structure of the Oort cloud, and \citet{not03} explains the anomalous acceleration as a 
manifestation of the cosmological constant. Additionally, a recent meeting at the University of 
Bremen\footnote{``The Pioneer Anomaly: Observations, Attempts at Explanation, Further Exploration,'' held at 
the University of Bremen, May 18-19, 2004, \url{http://www.zarm.uni-bremen.de/Pioneer}.}, a continuing 
series of meetings in Switzerland\footnote{``The Pioneer Explorer Collaboration: Investigation of the Pioneer 
Anomaly at ISSI'', \url{http://www.issi.unibe.ch/teams/Pioneer}.}, and a growing body of literature attest 
to the continued interest in the Pioneer Effect. For example, a recent preprint has discussed other external 
factors that could be related to the Pioneer Effect \citep{tur05}. In the guise of a problem set, this paper 
discusses a number of effects that could alternatively mask or explain the Pioneer Effect, including solar 
radiation momentum transfer, solar wind momentum transfer, solar corona electron density, Lorentz forces 
on a charged spacecraft, and clock instability and/or long term frequency stability. None of these effects 
are found to be of sufficient magnitude to explain the Pioneer Effect.

The bottom line is that the Pioneer Effect seems well-founded and has not been convincingly explained either in 
terms of known physics or engineering parameters of the spacecraft. Although spacecraft systematics 
remain the most likely explanation for the Pioneer Effect, its potential existence is of great interest for 
a variety of fundamental physical reasons.

Thus, the purpose of this paper is to assess the feasibility of using observations of minor planets 
to evaluate the gravitational field in the outer solar system and thereby explore the Pioneer Effect using 
precision astrometry. Although this methodology would have less temporal resolution and lower sensitivity to 
the magnitude of a detectable perturbing acceleration than would a spacecraft, this kind of observation program has 
the potential to be low in cost and to explore a possible perturbation effect along a number of vectors through the 
solar system which would otherwise require multiple spacecraft. Such an observation program could have profound 
effects on our understanding of the mass distribution in the outer solar system, and could also assist in 
discriminating between alternative gravitational theories such as MOND and classical gravity, as well as more 
exotic gravitational theories.

The remainder of this paper is organized into four sections. Section 2 describes the methodology used to 
evaluate the Pioneer Effect and the models used in the calculations. Section 3 addresses our results in 
terms of observational quantities that might be used to illuminate the mass distribution in the outer 
solar system. Section 4 discusses the results and addresses observational aspects of actually carrying
out the program outlined in this paper. Finally, Section 5 draws conclusions.

\section{METHODOLOGY AND MODELS}
The approach taken in this paper is twofold. First, we model the orbits of bodies in the outer solar system by means of 
Newtonian gravity and two-body, planar motion. We consider the Sun's field to be spherically symmetric 
and ignore the gravitational perturbations due to planets and the larger asteroids on the motion of the 
bodies of interest. While the importance of these effects is recognized, especially during the conduct of 
precision astrometry, they do not need to be considered in detail in the beginning of assessing the feasibility
of our approach. Initially, we only explore the perturbative effects of the Pioneer Effect on orbits. Since 
these effects, to first order, are linearly independent of other orbital perturbations (e.g., perturbations 
due to planets), the latter can be safely ignored. Other perturbations, for example the anisotropic thermal 
radiation giving rise to the Yarkovsky Effect, cometary nongravitational forces, General Relativity, and the 
Pioneer Effect will be addressed as appropriate. 

The general approach used in this phase of our analysis is to model the motion of bodies of interest subject 
to the perturbing forces of interest and to compare the heliocentric angular differences between the bodies 
in the different cases. The time evolution of these angular differences is considered along with the distances 
of the bodies and the precision with which their orbits are known to determine whether or not the effects of 
the perturbations can be detected and whether they can shed any illumination on the nature of the mass 
distribution in the outer solar system.

The approach outlined above is used for sample selection and provides a vehicle for first order exploration of 
the effects that might be observed if the Pioneer Effect were real. This approach would be complete in an 
ideal world, but we must perform our science in a messier reality, necessitating the second phase of our approach. 
Because the motions of minor planets are complex, we need to carry our analysis one step further. Since unavoidable 
observational errors and the motions of many perturbing bodies influence the dynamics of minor planets, the general 
approach towards understanding their motion involves determining orbital elements in such a way as to minimize 
discrepancies between observation and prediction. In our context, the problem is that adjustment of orbital parameters 
might allow motion perturbed by the Pioneer Effect to be masked completely and might make the Pioneer Effect 
unobservable in practice. For example, a change in orbital eccentricity might allow motion perturbed by the Pioneer 
Effect to be explained completely in terms of a non-Pioneer model. In this case, there would be no way to show 
the existence of the Pioneer Effect because the effect would be concealed beneath the variation resulting from 
measurement errors. 

In order to address these issues, we use the OrbFit software package \citep{mil99}. This program, freely 
available on the internet, uses observational data and data on the dynamics of the solar system to determine orbital 
elements and predict ephemerides for minor bodies. This tool is used in an extensive analysis of one of our 
candidate asteroids, (5335) Damocles. The primary purpose of this phase of the analysis is to demonstrate that the dynamic 
effects due to the Pioneer Effect are, in fact, measurable through observations. To that end, we added a simple option 
in OrbFit to include the force arising from the Pioneer Effect. By comparing hypothetical orbits with and without the 
Pioneer Effect, we explore the expected effect of such a perturbation on asteroid orbits and examine when this 
effect can be detected astrometrically.

Finally, we need to comment on the working definition of the Pioneer Effect used throughout this paper. Since the 
primary intent of this paper is to investigate the dynamical consequences of the Pioneer Effect, which apparently 
begin about 20 AU from the Sun, for simplicity, and because there are no data showing a more gradual onset of the 
Pioneer Effect, we will assume the anomalous acceleration of the Pioneer Effect begins abruptly at 20 AU. 

We recognize that this is a simplistic model of the Pioneer Effect. Alternative mechanisms exist that cause the Pioneer 
Effect to vary with object mass, orbital eccentricity, radial distance, and other parameters of the motion. As further 
observations of minor planets become available, they can potentially be used to investigate various force models in order 
to explore all possibilities until either the Effect is ruled out or its origin is found. However, the current status of 
information on the Pioneer Effect dictates that this simple model be investigated first. Furthermore, a perturbation 
beginning more gradually closer to the Sun would be more easily detectable. Thus our assumptions represent the minimum 
plausible perturbation from this effect given the available data.

\subsection{Minor Planets}
Before discussing minor planets, a few words on major planets are in order. These objects have been observed for very 
long periods of time and a great deal of effort has been devoted to explaining their motion. Why is it that evidence 
of the Pioneer Effect has not been seen in their motion? The answer is twofold. First, the orbits of the inner planets 
are known with great precision, with elements derived from highly accurate data including superb ranging data from 
numerous spacecraft, and do not show any evidence of the Pioneer Effect. This fits with our definition of the Pioneer 
Effect as having no influence at relatively small distances from the Sun. On the other hand the ephemerides of the 
outer planets are almost entirely based on optical observations \citep{sta04} and are much less accurate than those 
for the inner planets. In fact, Neptune has not even completed one revolution about the Sun since the introduction of 
reasonably sophisticated measuring instruments (e.g., the impersonal micrometer) in 1911 \citep{sta04} and Pluto has not 
completed a revolution since its discovery. Modern astrometry can obtain angular positions with reasonable accuracy, 
but the mean motions of the outer planets and their radial distances are quite uncertain. Thus, any Pioneer Effect 
perturbation on these bodies would be masked by uncertainty in the orbital semimajor axis.

Given these uncertainties, we must conclude that the outer planets do not represent good candidates for astrometrically 
determining the reality of the Pioneer Effect. Pluto might present such an opportunity, but its orbital elements require 
refinement. Many other bodies that go beyond 20 AU from the Sun have low eccentricities and, by extension, similar distance 
uncertainties, making their use for our purposes problematic and further reducing the number of candidates available 
for consideration. 

As far as comets are concerned, there are surprisingly few whose orbits are both known sufficiently accurately to be of 
interest and whose aphelion distance is greater than our assumed Pioneer Effect cutoff. Additionally, because of their 
extended natures, it is difficult to unambiguously determine the center of light of the comet and even that is not 
necessarily indicative of the actual location of the nucleus. Thus, determining the precise location of comets so that 
their orbits can be determined accurately is quite difficult. 

Comets also suffer orbital perturbations due to Non-Gravitational Forces (NGF) that presumably result from outgassing. The 
NGF are parametrized in a standard way \citep{mar73} and departures from osculating orbital elements provide estimates of 
the values of acceleration experienced by the comet. These forces have a substantial effect on the orbital parameters of 
comets, amounting to as much as several days difference in the predicted time of perihelion passage, a difference comparable 
to that produced by the Pioneer Effect. This standard parametrization is of a statistical nature. Since there is a variation 
in NGF from orbit to orbit, presumably as the comet's ``dirty snowball'' nature changes with multiple passes by the Sun, it 
is difficult or impossible to predict the exact motion of comets throughout their orbits. 

Additionally, a quick review of comets meeting our requirements show that they are extremely faint and exceedingly difficult 
to observe. This is due to the large distances from the Sun that candidates are found, their small size, and low albedo. 
Fairly typical is comet 1P/Halley. Recent observations of 1P/Halley have detected the comet at a distance of over 28 AU and 
at a visual magnitude of more than 28 \citep{hai04}. These observations represent the greatest distance and the greatest 
magnitude of any comet observation. The three 8.2m Very Large Telescopes at the European Southern Observatory's Paranal 
site were used simultaneously for a total exposure of 32 284 seconds in order to achieve this remarkable feat. 

The current magnitudes of candidate comets, coupled with the difficulties associated with NGF, force us to conclude 
that comets do not currently provide a good opportunity for investigating the Pioneer Effect. However, continuing cometary 
observations such as the long term program at the ESO \citep{hai04}, might provide insight over time, especially as target 
comets approach aphelion. 

Nevertheless, another interesting aspect of comets is that these recent observations of 1P/Halley show it to be approximately 
1.4 seconds of arc behind its ephermeris position. Although differences on the order of a second of arc potentially can be 
understood in terms of errors in the astrometry, the idea that perturbations due to the Pioneer Effect contribute is worthy 
of further investigation. 

Since low-eccentricity minor planets and comets are not suitable for the purposes of this paper, we hereafter restrict our 
attention to high-eccentricity objects whose orbits carry them sufficiently far from the Sun to be exposed to the Pioneer 
Effect as defined earlier. For brevity, in what follows we will use the term ``asteroid'' to indicate ``unusual'' minor 
planets, Trans-Neptunian Objects (TNOs), and Centaurs, which can be characterized by relatively large semimajor axes and 
relatively high eccentricities.

\subsection{Asteroid Sample Selection}
There are surprisingly few asteroids whose orbits are both sufficiently well known to be of interest and whose aphelion 
distance is greater than our assumed Pioneer Effect cutoff. Since asteroids are ``dead'' in an outgassing sense, they do 
not suffer NGFs as do comets. However, they do experience the Yarkovsky Effect, which is an acceleration resulting from 
anisotropic thermal radiation, and general relativistic perihelion precession. These effects will be addressed later, but 
using a selection criterion that asteroids have an aphelion distance greater than 20 AU, Horizons \citep{gio96} provided 
a candidate list of 985 asteroids. This selection criterion provides candidate objects that pass far enough from the 
Sun to be subject to the Pioneer Effect. In order that the asteroids approach closely enough to the Sun to enable them 
to be observed and their orbital elements to be determined with sufficient precision, we also excluded asteroids whose 
perihelion distances were greater than 20 AU. Further, in order that a reasonably significant part of an orbit be observed 
in a realistically short period of time, asteroids whose period of revolution exceeded 200 years were also excluded. 
Finally, asteroids with eccentricities less than 0.6 were excluded.

The eccentricity criterion deserves further explanation. If we consider a constant radial perturbation 
applied to a Keplerian orbit, Lagrange's planetary equations (in the Gaussian form) provide for a nonzero 
time rate of change in eccentricity, semimajor axis, mean motion, and argument of perihelion \citep{dan88}. 
If these rates are normalized by common factors, the normalized rate of change in eccentricity, argument 
of perihelion, and mean motion are smaller than that of semimajor axis by a factor at least as large as 
the semimajor axis. The only exception to this is for very small values of eccentricity, where the argument 
of perihelion can change quite rapidly. This can be understood by realizing that the primary manifestation 
of the Pioneer Effect lies in causing the orbit to precess. Considering a nearly circular orbit, a very 
slight precession can lead to a large angular change in perihelion position. It is very difficult to 
accurately determine a complete set of orbital elements of such an object; thus, we preferentially choose to consider 
more eccentric orbits, specifically those with eccentricity greater than 0.6. These bodies are selected because of 
the much larger changes in orbital elements associated with those objects than those with more modest eccentricities.

Table 1 shows selected orbital elements of the resulting list of 15 candidate asteroids with orbital 
geometry satisfying these criteria. 

\placetable{tbl-1}

\section{RESULTS}

Since asteroids are dynamically ``dead'' and do not outgas, and don't exhibit nongravitational accelerations as do many 
comets, they generally behave in a much more sedate and predictable way. Additionally, since they do not display comae, 
they are point sources and it is easy to unambiguously locate their positions. However, in regions where it is postulated, 
the magnitude of the Pioneer Effect is significantly less than the acceleration due to the Sun's gravity. At a distance 
of 20 AU, the gravitational acceleration due to the Sun is approximately 1.5$\times$10$^{-3}$ cm sec$^{-2}$, compared 
with the Pioneer Effect acceleration of 8.74$\times$×10$^{-8}$ cm sec$^{-2}$. Thus, because of its small magnitude, a 
number of external factors might contribute to or explain the Pioneer Effect.

One such phenomenon is the Yarkovsky Effect, which is a anisotropic reaction force associated with infrared reradiation 
of absorbed solar radiation. In the typical treatment, the Yarkovsky Effect is much more important for small bodies that 
are regolith-free than for larger objects, or those possessing a thermally insulating layer of regolith. 

The Yarkovsky Effect is generally considered to be of two forms, the ``diurnal'' effect occurs when the rotation of the 
body about its axis causes reradiation to occur at a different ``time of day'' than when the solar radiation was absorbed. 
The ``seasonal'' Yarkovsky Effect occurs, for example, when the rotation period of the object about its axis is much shorter 
than the orbital period. When this occurs, the ``diurnal'' thrust averages to zero, while the reradiation occurs at different 
times in the body's orbit about the Sun \citep{spi01}. The diurnal effect can either expand or contract orbits; the seasonal 
effect always shrinks orbits \citep{rub95}. Detailed expositions on the magnitude of the Yarkovsky Effect are available in 
the literature (e.g., \citet{rub95}), but a simple estimate shows that the acceleration due to the Yarkovsky Effect is 
inversely proportional to the asteroid's density and radius, and inversely proportional to the square of the distance from the 
Sun. For an asteroid 20 AU from the Sun, with a radius of 200 km and a density of 2 g cm$^{-3}$, the Yarkovsky acceleration is 
more than seven orders of magnitude smaller than the Pioneer Effect acceleration and even more negligible compared to the 
acceleration due to the Sun at that distance.

Another external source that might explain or at least contribute to the Pioneer Effect is the general relativistic orbit 
precession. According the standard Parametrized Post-Newtonian (PPN) approximation in general relativity (e.g., \citet{mis73}), 
the greatest value of the orbital period change due to general relativity for all our asteroid candidates is of the order of 
seconds. The corresponding minimum orbital period change due to the Pioneer Effect is of the order of five hours. Thus, as is 
normally expected from a general relativistic effect, the magnitude of the PPN perihelion precession is negligible in comparison 
with that due to the Pioneer Effect. 

Thus, there is a sample of asteroids that provide a clean and unambiguous vehicle for exploring the gravitational field in the 
outer solar system. Their inert dynamical nature, coupled with their relatively high visibility from Earth allows their motion 
to be characterized and predicted with assurance, and allows deviations from predicted motion to be measured readily in reasonable 
periods of time.

\section{DISCUSSION}

\subsection{Asteroid Dynamics}
The previous discussion of asteroid orbits is similar to a ``kinematic'' approach, wherein we explore features 
of the motion without regard to its specifics. However, a ``dynamic'' assessment, incorporating the current 
location of asteroid candidates in their orbits is now of interest in order to determine whether the Pioneer 
Effect is observationally detectable. Of the 15 asteroid candidates, only two are currently outside the 20 AU 
boundary, with one moving outward toward aphelion and the other moving inward. Seven are currently beyond 10 AU 
and are moving outward, while one is that far away and is moving inward. The remaining five closer asteroids 
are all currently moving outward.

If the Pioneer Effect is real, the asteroids that are currently beyond 20 AU have already had their 
positions perturbed relative to their ephemerides without the Pioneer Effect. Of the candidate asteroids 
there are only two that fit this category. (5335) is currently at 20.8 AU and is barely into the Pioneer 
Effect region. 1995SN55 is currently at 38.4 AU and is past aphelion on its way back to the inner solar 
system. Predictions show that the former has not developed a measurable angular deviation in the short 
time it has been further than 20 AU from the Sun; the latter has been in that region for over 54 years 
and has deviated from an orbit unperturbed by the Pioneer Effect by about 30.5 seconds of arc. This 
level of angular deviation should certainly be observable. Table 2 provides data on the current positions 
of the candidate asteroids.

\placetable{tbl-2}

If the current positions of 1995SN55 is corrected for this initial discrepancy, and the unperturbed 
and Pioneer-perturbed orbit is made to coincide at the current epoch, we can plot the rate at which the 
angular deviation grows from the present. Thus, Figure 1 shows the heliocentric angular deviation of each 
asteroid from a starting point of 2005 April 1. The deviation shown is the ``Observed minus Calculated'' 
deviation with the ``calculated'' orbit being that perturbed by the Pioneer Effect. 

\placefigure{fig1}

Astrometry with current CCD techniques is routinely accurate to 0.3--0.5 seconds of arc for objects like our 
asteroid candidates. Thus, Figure 1 would seem to indicate that several asteroids are good candidates for 
observations to measure the Pioneer Effect. However, to consider observational constraints on the candidate 
asteroids, we must consider two additional issues in addition to orbital geometry: 
\begin{itemize}
\item First, that the current ephemeris uncertainty is low enough that observation without extended search 
is likely;
\item Second, that the asteroid is large enough or bright enough to allow a reasonable expectation of 
observation over the majority of its orbit.
\end{itemize}
A figure of merit for the current ephemeris uncertainty has been developed and is provided by the 
Minor Planet Center (MPC) in the orbital elements as the $U$ parameter. The MPC defines the $U$ parameter 
``in order to quantify the uncertainty in a perturbed orbital solution for a minor planet in a concise 
fashion.'' $U$ is an integer ranging from zero to nine, corresponding to the uncertainty per decade 
along the Line Of Variance (LOV) of the object's orbit. Zero indicates a very small uncertainty and 
nine an extremely large uncertainty in the orbit\footnote{Further explanation can be found at 
\url{http://cfa-www.harvard.edu/iau/info/UValue.html}.}. 

An examination of the last observation history files at the MPC for Trans-Neptunian and Kuiper Belt 
Objects shows that of the sites currently submitting astrometric measurements, a visual magnitude limit 
between 26 and 27 appears to be the current capability for ground-based observation, with the very largest
instruments being capable of reaching a magnitude of 28. Figure 3 shows the maximum observable distance as 
a function of absolute magnitude for several limiting magnitudes in this range. Also shown in this figure 
are the points corresponding to the candidate asteroids.

\begin{figure}
\plotone{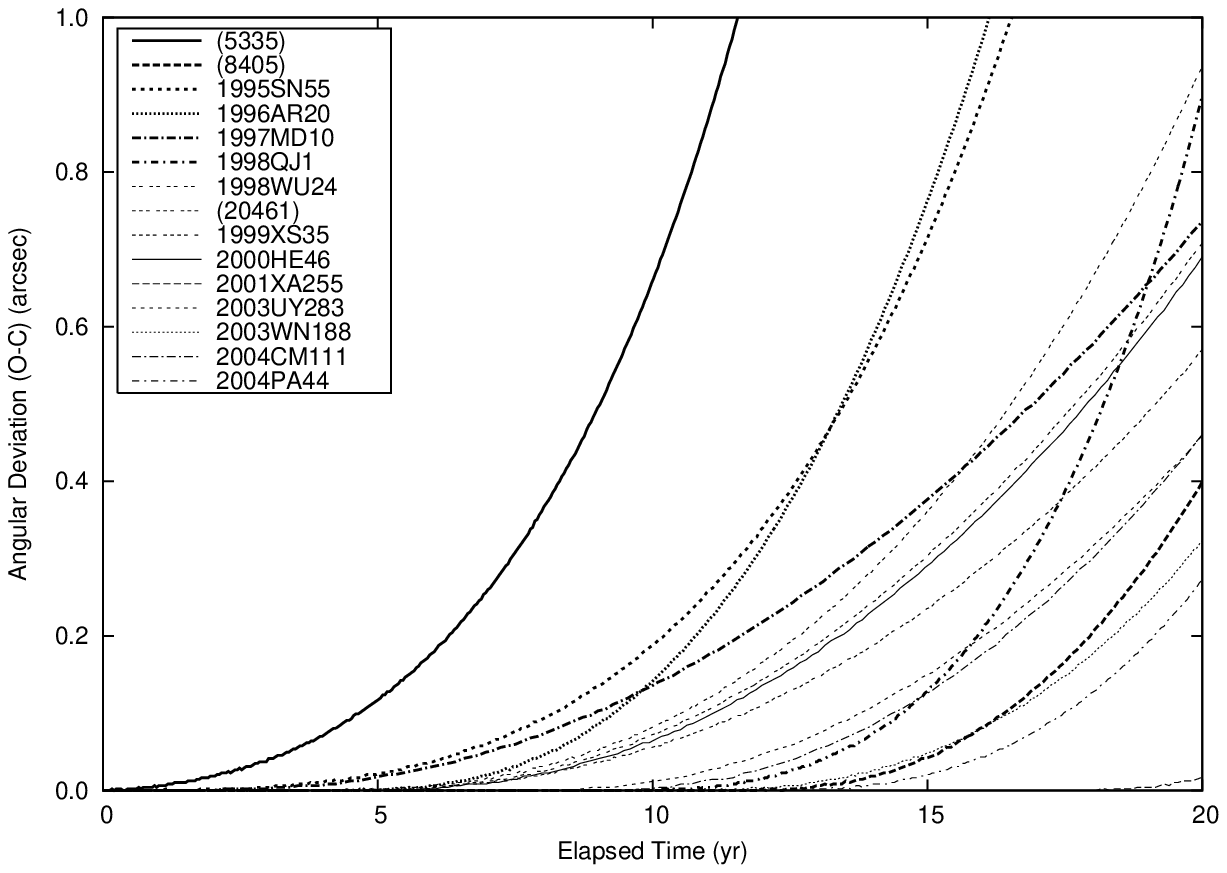}
\caption{Heliocentric angular deviation between Keplerian orbit for asteroid candidates
         and those perturbed by the Pioneer Effect. The elapsed time is that from
         2005 April 1.\label{fig2}}
\end{figure}

\placefigure{fig2}

From the data in Table 2 and Figure 2 it is not difficult to conclude that of the 15 known asteroids 
with suitable orbit geometry, only five have a size and brightness sufficient to allow observation over 
either a majority of their orbit or a significant period of observation of their orbit beyond 20 AU. These 
objects are (5335), (8405), 1995SN55, 1996AR20, and 2004PA44.

Thus, these asteroids should provide a mechanism for observing the gravitational field in the outer 
solar system and permit its use in investigating the Pioneer Effect and, in a broader context, the mass 
distribution in the outer solar system. Additionally, many of the other candidate asteroids could be 
observed in the near future, when they are not in the Pioneer Effect region, in order that their orbits 
be tied down with observations when they are close. This could be done in anticipation of continuing 
observations when they move further out and become subject to the Pioneer Effect.

\subsection{Observational Issues}
What observational issues are associated with using our candidate objects to investigate the Pioneer 
Effect? The analysis presented above would be quite complete in an ideal world without measurement errors. 
Of course, reality is messier. One must take observations that contain errors, and fit an orbital 
solution to them in some way as to minimize the discrepancies between prediction and observation. 
Generally, this is done using a description of the orbit (e.g., orbital elements) and minimizing the 
total square deviation of the orbital solution from observations in a least squares sense. Thus, exact 
orbit solutions are not available; rather, orbits with various uncertainties and different goodness of 
fit statistics are what results.

The dynamics of minor planets in the solar system are complicated. Not only do observational errors affect 
the outcome, but the motion of the planets and other perturbations in all their complexity impact the minor 
planet's motion as well. The problem arises because an adjustment of orbital parameters may allow a given 
set of observations to match a perturbed orbit. For example, a change in eccentricity might allow motion 
perturbed by the Pioneer Effect to be explained completely in terms of a non-Pioneer model. In this case, 
there would be no way to distinguish whether or not the Pioneer Effect existed because there would be no 
observational consequence associated with it. The question remaining, then, is whether the Pioneer Effect 
can be distinguished observationally given the uncertainties associated with orbit determination.

OrbFit software was used to investigate the motion of one of our candidate asteroids, (5335) Damocles, to determine 
if the Pioneer Effect could produce truly observable consequences in the motion of this object. Four cases were 
investigated. First, the existing observations, numbering 51\footnote{It should be noted that these observations 
represent the entirety of those available from the archives of the Minor Planet Center. We emphasize the necessity 
of using all available observations of the objects under consideration in order to obtain the best possible orbit 
characterization.} and occurring over two oppositions from 1991 February 18 to 1992 August 22 were used to determine 
orbital elements and ephemerides for (5335) when it was not subjected to the Pioneer Effect and otherwise. These 
real observations are all of high quality, and the astrometry is derived from the associated CCD images. The 
second set of predictions are associated with the 51 existing observations plus another four synthetic observations 
performed ``now,'' specifically 2005 June 1, 3, 15, and 17. Two sets of observations were synthesized, assuming 
normally distributed measured positional rms errors of one second of arc. The first set was based on ephemeris 
position predictions with the 51 real observations but without the Pioneer Effect, while the second set of 
observations was based upon ephemeris predictions using the real observations with the Pioneer Effect. Note that 
the orbital elements in the two cases were different as the synthetic observations giving rise to them are 
different. However, the procedure used was parallel between the two cases, differing only in the force model used.

OrbFit provides not only predictions of ephemeris position, but also estimates of the positional error on the sky 
associated with each prediction. The problem of determining these errors is nonlinear in nature and cannot be 
solved in general. Often, a linear approximation is used, and although OrbFit offers a semi-linear approximation 
that can be considerably more accurate than that provided by the linear approximation, investigation showed that 
for the magnitude of the angular differences considered here, the linear approximation is completely adequate 
\citep{mil99}.

The result of running OrbFit is that, for each case with and without additional synthetic observations and with 
and without the Pioneer Effect perturbation, we have an ephemeris showing position on the sky as a function of 
time along with the error estimates at each instant. The error estimates are given as one standard deviation 
error in a maximum direction (with an associated position angle), and the one standard deviation error in an 
orthogonal direction. Thus, equal probability loci form ellipses about the predicted position. 

We determine the the angular differences between the four cases. Associated with each positional difference is a 
direction, easily specified in terms of a position angle $\theta_{pos}$, and each position has an associated 
observational error ellipse. We are interested in the projection of this error in the direction of the angular 
difference between the predicted positions. If $E_{1}$ and $E_{2}$ are the semi-axes of the error ellipse (by 
construction $E_{1}$ is the greater of the two) and $\theta_{err}$ is the position angle of the major axis of 
the error ellipse, the magnitude of the error in the direction of the angular difference between the positions is
\begin{equation}
\sigma = { { E_{1} }
             \over
           { \sqrt{ 1 + [ { ( { E_{1} / E_{2} } ) } ^{2} - 1] \sin ^{2} { ( \theta_{pos} - \theta_{err} ) } } } }
\end{equation}
Statistically, we can test the hypothesis that the orbits are the same by considering the difference between the 
predicted positions and comparing this quantity with a confidence interval at the appropriate level of significance. 
The positional uncertainty used in this calculation is the square root of the sum of the squares of the independent 
errors associated with the two positions.

Figure 3 shows the results of using OrbFit to investigate these questions. The upper panel of the figure shows the 
angular difference between the without-Pioneer and with-Pioneer cases, and a 95 per cent confidence interval for the 
case with only the original 51 real observations being available. Each case has a one standard deviation error ellipse 
at the present time with semimajor and semiminor axes approximately 7 arcsec and 0.5 arcsec in size, respectively, and 
is oriented with the long axis having a position angle of approximately 12 degrees. Most of the predicted angular position 
difference is in the declination direction, parallel to the long axis of the error ellipse. Thus, the larger error ellipse 
dimension contributes most to the confidence interval. The hypothesis that the two cases are the same is rejected at the 
five per cent level if the 95 per cent confidence interval does not encompass zero. As can be seen from the figure, since 
the errors grow faster than the angular difference, this never occurs in the time interval shown in the figure, and likely 
for a considerable time thereafter.

\begin{figure}
\plotone{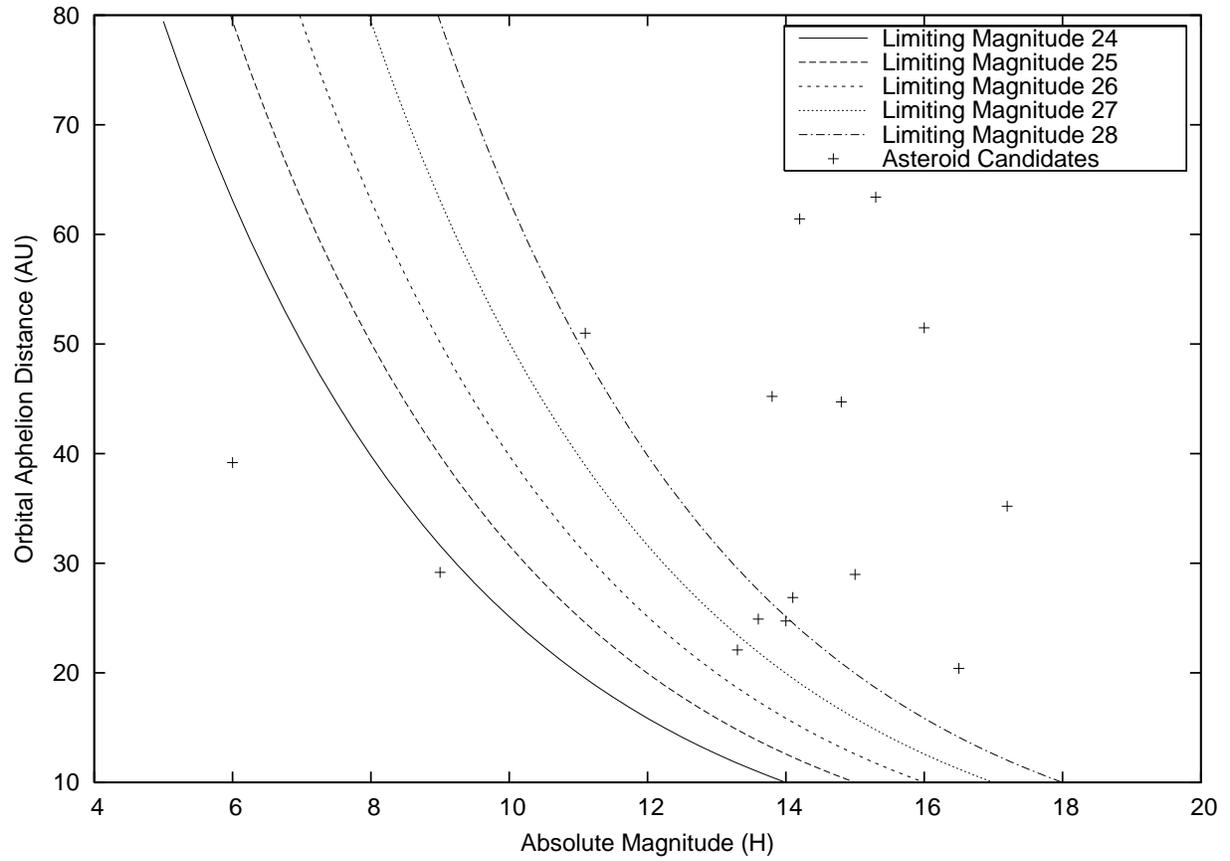}
\caption{Maximum distance observable as a function of object absolute visual magnitude for
         various limiting magnitudes. This figure was generated using the definition of visual
         magnitude simplified to $m_{V}=H+5 \log {r^{2} }$. The points shown represent the 15 
         candidate asteroids.
         \label{fig3}}
\end{figure}

\placefigure{fig3}

The lower panel of Figure 3 shows corresponding results when four additional synthetic observations in June 2005 are made 
as described above. One notes that the starting point of the two panels is different. This is due to the fact that the 
synthetic observations in June 2005 are different for the without-- and with-Pioneer Effect cases since Damocles would have 
been in the Pioneer Effect region for some time when the synthetic observations are ``conducted.'' Once the orbit is adjusted, 
this results in different orbital elements and ephemerides. As expected, the additional observations shrink the total error 
ellipse substantially. The synthetic observations lead to an error ellipse with semimajor and semiminor axes approximately 
0.5 arcsec by 0.4 arcsec in size, with the long axis having a position angle of about 12 degrees. In this case, most of the 
position difference remains in the declination direction, parallel to the long axis of the error ellipse. However, since the 
size of the error ellipse is substantially reduced, especially in this direction, the size of the confidence interval is 
substantially reduced as well. Geometric effects relating to the positions of the Earth and (5335) make the orbital errors 
time variable. With observations occuring now, the hypothesis that the with-- and without Pioneer Effect cases are the same 
is rejected at the five per cent level for the first time at about MJD 56 958 or 2014 October 27. After that time, such 
determinations occur more and more frequently as the predicted orbits grow further and further apart. Thus, observations of 
(5335) performed now could refine its orbit and allow, within a few years, a relatively unambiguous determination of whether 
the Pioneer Effect has influenced the motion of the asteroid.

In all the cases discussed above, the rms residual is 0.7 arcsec, indicating a well-characterized orbit. Interestingly, 
if the synthesized observations are switched and the no-Pioneer Effect synthetic observations are used without the 
Pioneer perturbation and vice versa, the residuals do not change from this value. This is more a comment on the original 
51 observations being performed in a fairly short period of time, with the additional synthetic observations being 
temporally separated by a fairly large interval than anything else.

In the analysis just presented, synthetic observations were performed ``now'' (June 2005) to ``pin down'' Damocles' orbital 
parameters. Then, the evolution of the orbit in time was compared for the Pioneer-perturbed and unperturbed cases. As time 
progressed, this involved implicit additional observations for both cases, but without bringing those observations into the 
orbital element calculation. What happens if more observations are performed and the additional observations are used to 
fit Damocles' orbit? 

To address this question, two sets of  ephemerides were generated using the 1991-1992 actual observations of Damocles as a base. 
The first set included only normal orbital forces, while the second set contained the additional perturbations expected from the 
Pioneeer Effect. From these data, sets of synthetic observations were created which included a Gaussian astrometric uncertainty 
of $0.3$ arcsec in both right ascension and declination, appropriate for current high quality astrometry. We assumed a 90 day 
observation period every year starting in 2006. During each year, we used a total of three positions from May, June and July as 
the new astrometric measurements. The orbit was then analyzed with the two versions of OrbFit, and the residual was tabulated for 
the period from 2006 to 2026, with each new fit including all the previous real and synthetic observations up to that time. The 
result of this experiment is the trend in the total residual of the orbital fit as a function of time for the orbits with and without 
the Pioneer Effect. There are four cases, consisting of the combinations of observations synthesized with-- and without the Pioneer 
Effect perturbation as Damocles moves under the influence of the Pioneer-perturbed and unperturbed gravitational force. In the 
two cases where the synthetic observations match the force model, we would expect the residual to slowly decrease as additional 
observations are added. In the two cases where the genesis of the synthetic observations does not match the force model, the model 
fit residual should increase as additional observations are added.

To gain understanding of the robustness of our fit, we repeated this experiment 100 times using different astrometric observational 
errors. Each run had the same $0.3$ arcsec deviation for each observation but used a different normally distributed random value 
for the astrometric error. The net result of this experiment is shown in Figure 4. In the top panel, the residuals are shown 
for the orbits generated with observations produced with and without the Pioneer effect, but analyzed without the Pioneer Effect 
force perturbation. The error bars represent a one standard deviation variation from the average of our ensemble of 100 runs. By 
2024, the error bars separate as the residual from fitting the orbit whose synthetic observations included the Pioneer Effect 
increases. This separation will continue to grow as the number of observations increases.

We also fit the synthetic observations generated with the standard and Pioneer Effect perturbed models to a modified version of 
OrbFit that included the Pioneer effect in its force model. In this experimental case, we test the opposite hypothesis: ``How 
well does the Pioneer Effect fit the observations?'' instead of ``How well does a normal orbit fit the observations?'' The results 
of this fit in the lower panel of Figure 4. As expected, the residual monotonically decreases with the orbit that includes the 
Pioneer Effect when using this code. However, the residual for the orbit that does not include the Pioneer Effect deviates from 
this slow decrease earlier than it did in the previous case. Although we have no specific cause for the apparent asymmetry between 
the two cases, it is not unexpected given the nonlinear nature of these calculations. However, fitting an orbit to a code that 
includes the Pioneer Effect would likely give a more rapid and robust result to the question of the existence of this effect 
than fitting an orbit to a code that includes only the standard gravitational perturbations.

The result of these numerical experiments is to show that a modest observational program with only a few observations a year 
should be able to determine if the Pioneer effect is real in less than 20 years. With better accuracy and more frequent 
observations, this time could be substantially decreased.

\begin{figure}
\epsscale{0.60}
\plotone{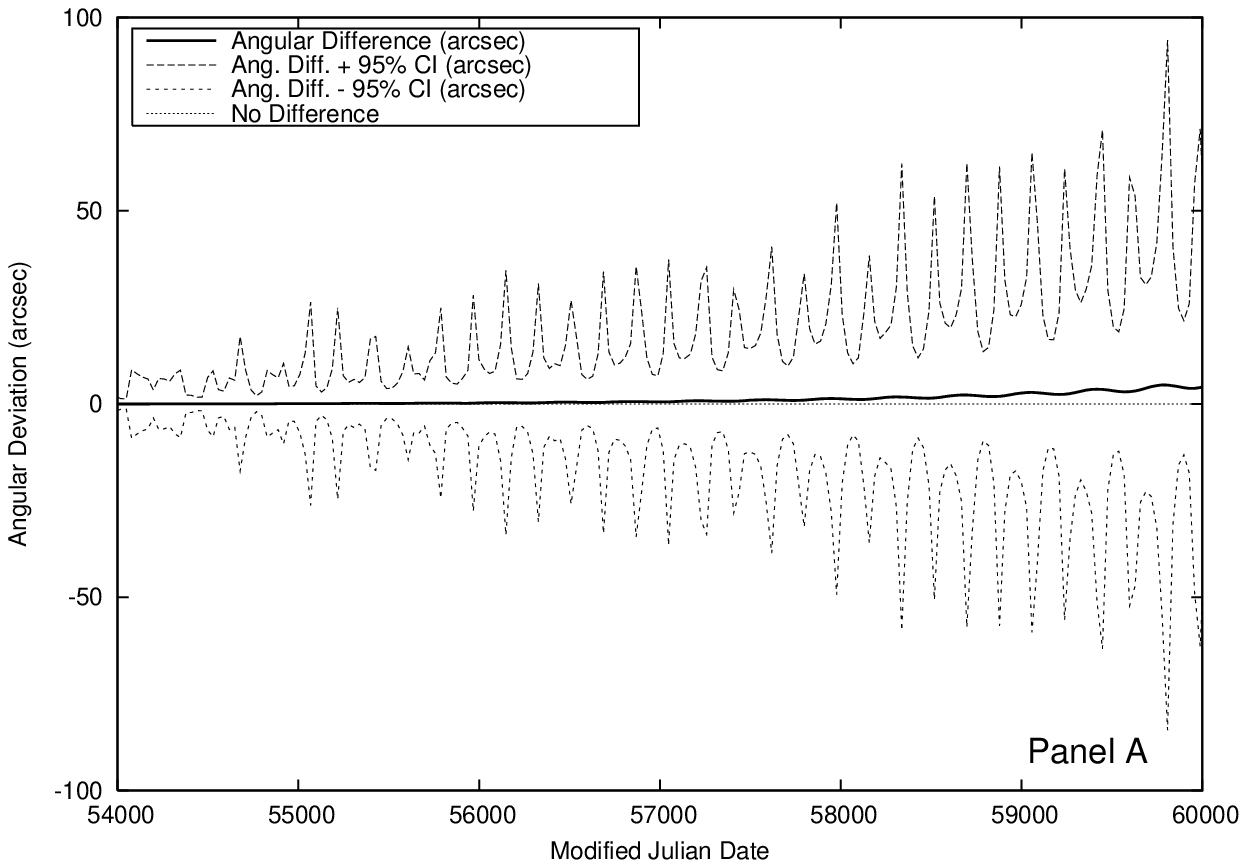}
\plotone{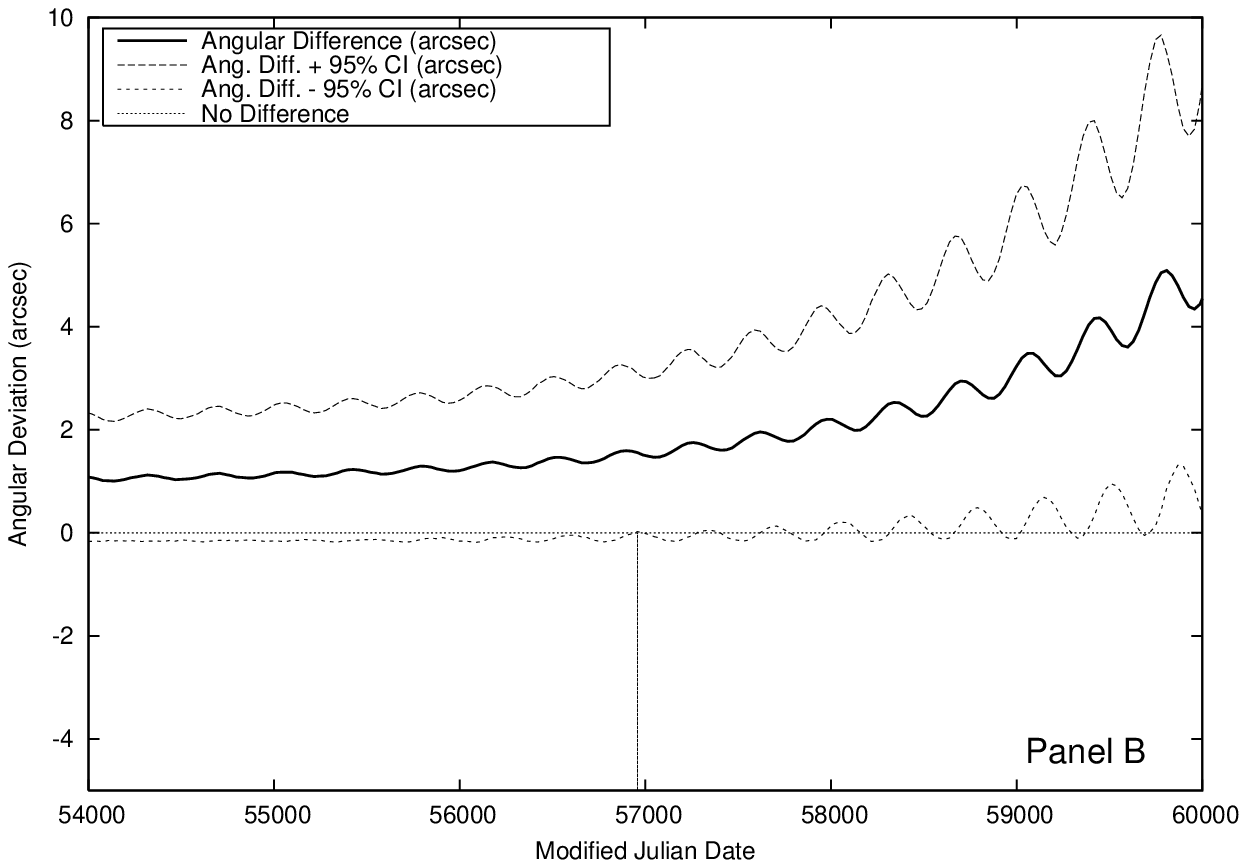}
\caption{Angular differences between positions of (5335) Damocles as a function of time with-- and without a Pioneer 
Effect perturbation. The horizontal axis runs from 2005 Sep 21 to 2023 Feb 24. The solid line in each panel shows the 
expected observational difference in position in the two cases. The upper and lower dashed lines in each panel represent 
a 95 percent confidence interval about the calculated difference in position. The upper panel (A) shows the results with only 
the currently available 51 real observations, but with the orbit adjusted to fit the perturbed and unperturbed cases. The 
dotted lines in each panel show a difference of zero between the two cases. Without additional observations, the hypothesis 
that the orbits are different can never be rejected at the five per cent level since the 95 per cent confidence interval 
always encompasses zero. The lower panel (B) shows similar results when the actual observations are combined with synthesized 
observations ``performed'' in June 2005. The hypothesis that the orbits are the same is rejected at the five percent level 
after about end-October 2014 (MJD 56 958).\label{fig4}}
\end{figure}

\placefigure{fig4}

Damocles' current position is 20.8 AU away from the Sun, moving outwards. It is currently at a visual magnitude of 26.8. 
Observing this object now, while challenging, is not impossible. An instrument in the four meter class would be sufficient 
to acquire the required astrometry. The well-characterized orbit of this object should allow it to be observed without a 
significant search, and current observations could assist in making a significant statement about the Pioneer Effect. At 
aphelion, Damocles' visual magnitude should be slightly greater than 27, allowing the object to be observed over the entirety 
of its orbit. It will be in the Pioneer Effect region, as defined in this paper, until late November 2018. Thus, there is 
much time to observe this object and to use its motion for exploring the Pioneer Effect.

Of our other four candidate asteroids, two are in well-characterized orbits. The orbit of (8405) Asbolus has been 
well-observed over ten oppositions from 1995 through 2004, and has rms residuals of 0.6 arcsec. Its maximum visual 
magnitude at aphelion is less than 24, allowing its motion to be observed over its entire orbit with reasonably available 
instruments. It is currently not in the Pioneer Effect region, is still moving outward, and will not cross 20 AU 
until about 2016 June 8. This object presents an opportunity for further orbital characterization and possible exploration 
of the onset of the Pioneer Effect if it exists.

2004PA44 has been observed over three oppositions from 2002 through 2004. It has rms residuals of less than 0.4 arcsec. 
At aphelion, its visual magnitude is approximately 26.6, allowing it also to be observed over its entire orbit. It is also 
not currently in the Pioneer region, is moving outward, and will not enter the Pioneer region until the end of December 2016. 
As with Asbolus, 2004PA44 provides an opportunity for further orbital refinement and potential investigation of the beginning 
of the Pioneer region.

Our remaining two candidates can only be considered lost. It would require a significant search effort to reacquire these 
objects because of their short observation arcs. 1996AR20 was observed for a short period in 1996, is currently moving outward, 
and is not expected to reach the Pioneer region until about April 2009. Its visual magnitude at aphelion, nearly 28, would make 
this a very challenging target for observation over the whole of its orbit. However, if it can be located, it too could offer 
an opportunity for orbit refinement and further Pioneer Effect investigation.

1995SN55 is also lost, however, in many ways it is the most intriguing of our candidate asteroids. It was observed over a 
short arc in 1995 and has not been observed since. It is currently over 38 AU away from the Sun, well into the Pioneer region, 
moving sunward, and is just past aphelion. However, its large size makes its current visual magnitude only slightly greater 
than 22. If this object could be located, over a span of years it would offer an excellent opportunity to measure the 
gravitational field in the outer solar system, as well as determining whether the Pioneer Effect exists or not.

All of these objects possess excellent, high quality CCD astrometry, albeit in many cases not enough either in number of 
observations or in temporal currency. As observed above, all the objects could be observed over their entire orbits and, over 
time, could provide an excellent vehicle for exploring gravity in the outer solar system whether that involves the mass 
distribution in those regions, or more exotic physics.

\section{CONCLUSIONS}

The purpose of this paper is to assess the feasibility of using observations of minor planets to evaluate the gravitational 
field in the outer solar system and thereby explore the Pioneer Effect by means of  precision astrometry. 

If a method of measuring the Pioneer Effect was available it might serve, once and for all, to either support or refute its 
existence as a real phenomenon. We show that asteroids can fill this role. These bodies are useful for this purpose because 
they have a large mass and are large and bright enough to observe for satisfactorily long intervals. Our analysis clearly shows 
that observations could determine whether or not the Pioneer Effect exists, and demonstrates that the residuals of orbital fits 
grow if continuing observations are conducted and are modeled with the erroneous force model. We further demonstrate that the 
Pioneer Effect could be confirmed or refuted by means of a sustained observation campaign against properly chosen asteroids. 
These observations can be conducted with modest allocations of telescope time, and would provide a definitive answer to the 
question within the next twenty years. 

Whether or not the Effect was substantiated, astrometry of asteroids can be used to measure the gravitational field in the 
outer solar system. Depending upon the number and type of the measurements, it might even be possible to break the degeneracy 
in the alternative predictions of different possible explanations for the Effect or differentiate between alternative 
gravitational theories. This is a worthwhile program in its own right, and observations of (5338), (8405), and 1995SN55 would 
be particularly helpful for this purpose. 

The proposed method is not without weaknesses, however. The first weakness of our approach is that it has less temporal 
resolution and lower sensitivity to perturbations than would a spacecraft, dedicated or otherwise. However, the proposed 
observation program is low in cost and can explore possible perturbation effects along a number of different vectors through 
the solar system, which would require multiple spacecraft. Such an observation program could have profound effects on our 
understanding of the gravitational field and implied mass distribution in the outer solar system, and could also assist 
in discriminating between alternative gravitational theories such as MOND and classical gravity, as well as more exotic 
gravitational theories.

A second weakness is that our adopted model of the Pioneer Effect is very simple. Since the primary intent of this paper is to 
investigate the dynamical consequences of the Pioneer Effect, which apparently begin at large heliocentric distances, for 
simplicity, and because there are no current data supporting a more gradual onset of the Pioneer Effect, we assumed the 
anomalous acceleration of the Pioneer Effect begins abruptly at 20 AU. This assumption, however, does not compromise the 
methodology. It only places limits on the number of asteroids to consider as observational candidates. We recognize this as 
a simplistic model of the Effect. Alternative mechanisms exist that cause the Pioneer Effect to vary with object mass, 
orbital eccentricity, radial distance, and a number of other parameters of the motion. As further observations of minor 
planets become available, they can potentially be used to investigate various force models in order to explore all possibilities 
until either the Effect is ruled out or its origin is found. However, the current status of information on the Pioneer Effect 
dictates that this simple model be investigated first.

Despite the limitations of the use of asteroids in the roles addressed here, it remains true that there are very few intermediate 
range tests of gravity at the multiple AU distance scale. Comets experience reaction forces due to outgassing, and the outer 
planets move very slowly and cover only some of the region of interest. Spacecraft like Pioneer are expensive, as well as being 
tiny, fragile things that outgas, get pushed about by solar winds, and suffer reaction forces due to their radio transmissions 
and power sources. Most newer spacecraft improve their guidance capabilities by conducting mid-course corrections, leading to 
more motion variation and greater difficulty in discerning the small perturbations. With all the limitations of the proposed 
method, there is nothing quite as useful as a big, unwieldy, dynamically dead chunk of rock for investigating small variations 
in Newton's Laws or the mass distribution in the outer solar system.

\acknowledgements
The authors wish to acknowledge the Minor Planet Center for observational data on (5335) Damocles, available through their 
Extended Computer Service\footnote{\url{http://cfa-www.harvard.edu/iau/services/ECS.html}}. Additionally, the excellent software 
packages developed and maintained by the OrbFit Consortium\footnote{\url{http://newton.dm.unipi.it/orbfit}} and the FindOrb 
program developed by Bill Gray and Project Pluto\footnote{\url{http://www.projectpluto.com/find\_orb.htm}} allowed orbital 
calculations to be performed with the requisite precision. Both programs made use of JPL's DE405 ephemeris 
data\footnote{\url{http://ssd.jpl.nasa.gov/eph\_info.html}} to describe the dynamics of the solar system. Finally, the authors 
would like to thank the anonymous referee for his thoughtful and useful comments that resulted in significant improvements to this paper.

\clearpage

\clearpage

\clearpage

\clearpage
\begin{figure}
\epsscale{0.60}
\plotone{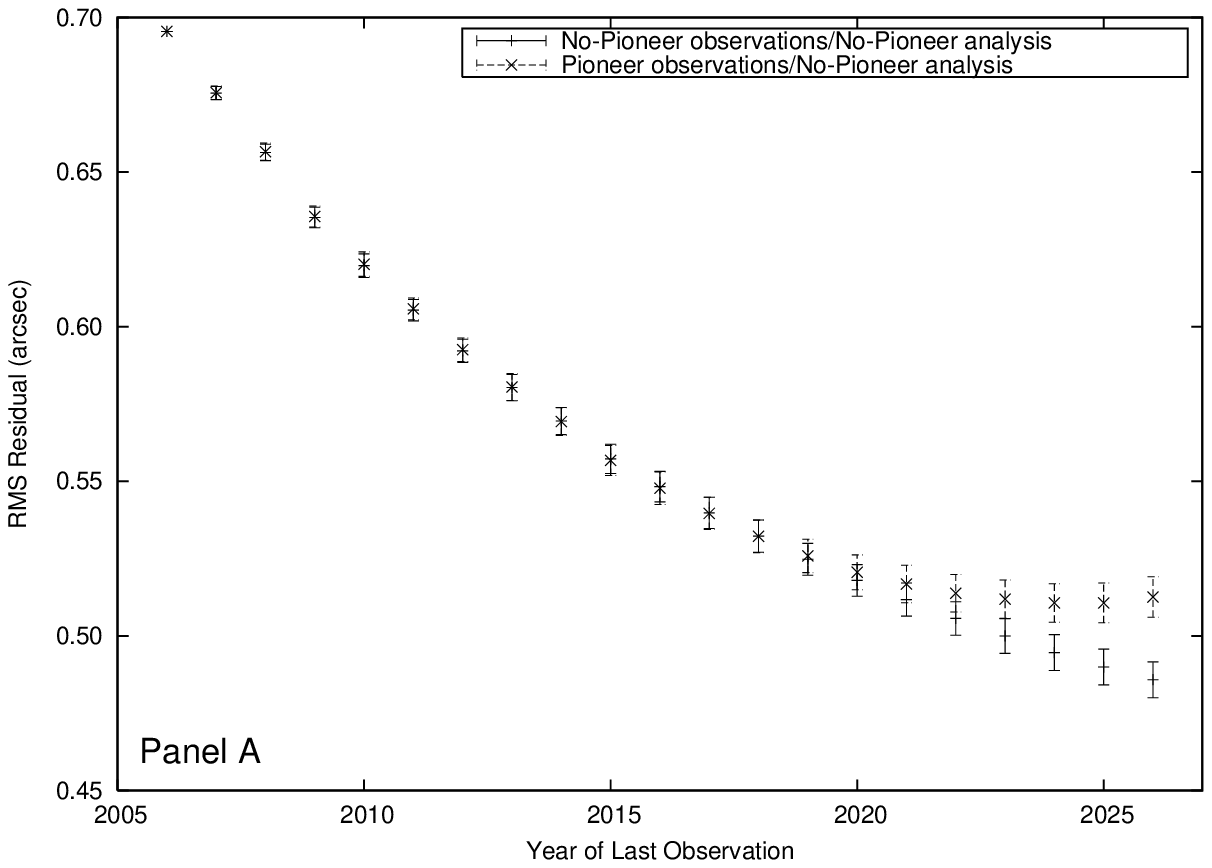}
\plotone{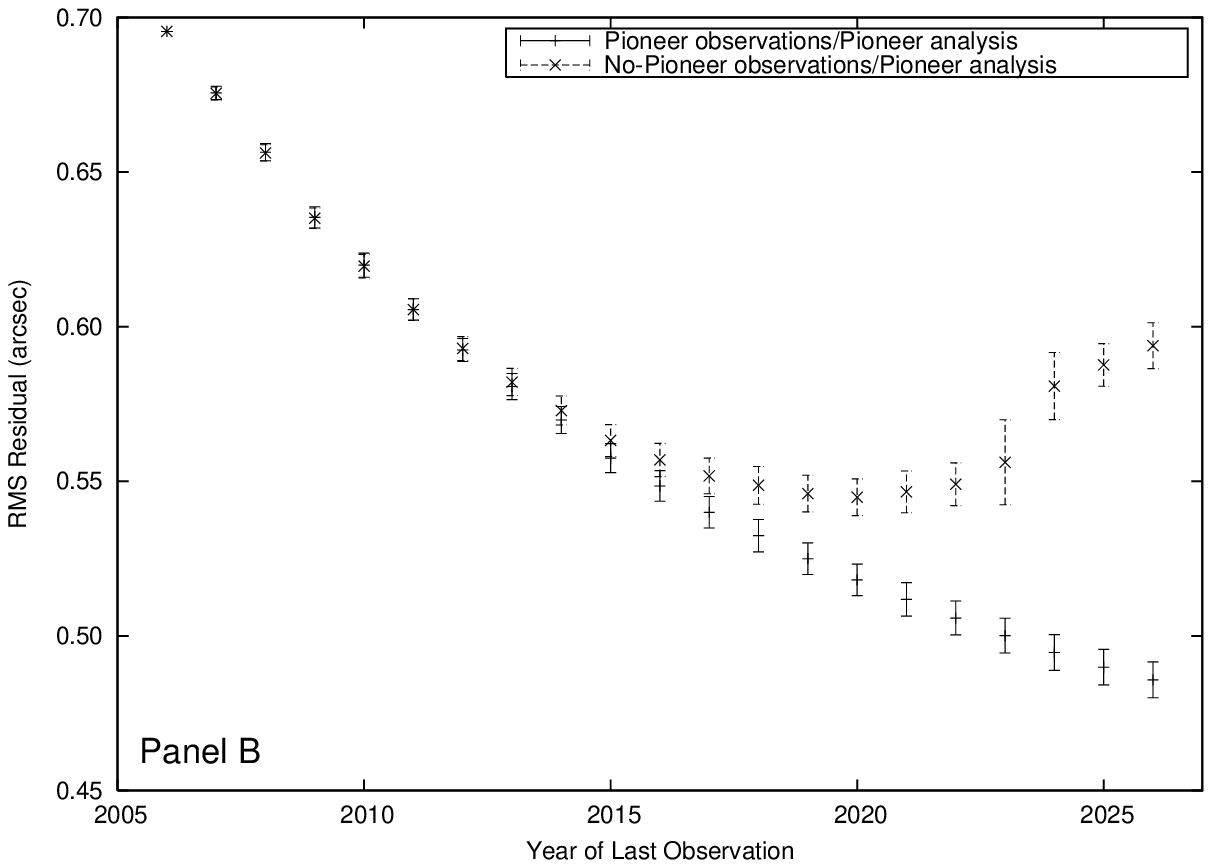}
\caption{Orbital fit rms residuals for Damocles as synthetic observations are added. The first term in the legend describes 
the case used to generate the synthetic observations; the second term describes the force model used to conduct the analysis 
of the observations. Sets of three synthetic observations were added annually. After each set of observations, the orbital fit 
was updated. There are four cases, consisting of the combinations of observations synthesized with-- and without the Pioneer 
Effect perturbation as Damocles moves under the influence of the Pioneer-perturbed and the unperturbed gravitational force. 
The top panel (A) shows the results of the case where synthetic observations are generated for both cases, but the motions 
are analyzed using a force model that does not include the Pioneer Effect. The bottom panel (B) shows the other two combinations, 
where the same synthetic observations are analyzed with the Pioneer Effect gravitational perturbation. The residuals initially 
decline, but those of the mismatched observations and force models eventually begin to grow larger, while the residuals of the 
matched observations and force models cases continue to decline.\label{fig5}}
\end{figure}

\clearpage
\begin{deluxetable}{lccccc}
\tabletypesize{\scriptsize}
\tablecaption{Orbital parameters of asteroids susceptible to the Pioneer Effect.\label{tbl-1}}
\tablewidth{0pt}
\tablehead{
\colhead{Asteroid} & \colhead{$a$} & \colhead{$e$} & \colhead{$T$} & 
     \colhead{$Q$} & \colhead{$A$}
}
\startdata
(5335)    & 11.837 & 0.866989 &  40.73 & 1.574 & 22.099 \\
(8405)    & 17.999 & 0.620309 &  76.36 & 6.834 & 29.164 \\
1995SN55  & 23.564 & 0.663131 & 114.39 & 7.938 & 39.190 \\
1996AR20  & 15.197 & 0.627202 &  59.25 & 5.666 & 24.729 \\
1997MD10  & 26.509 & 0.941736 & 136.49 & 1.545 & 51.474 \\
1998QJ1   & 11.255 & 0.813052 &  37.76 & 2.104 & 20.405 \\
1998WU24  & 15.201 & 0.906005 &  59.27 & 1.429 & 28.973 \\
(20461)   & 23.803 & 0.899499 & 116.13 & 2.392 & 45.213 \\
1999XS35  & 18.079 & 0.947578 &  76.87 & 0.948 & 35.210 \\
2000HE46  & 23.540 & 0.899577 & 114.22 & 2.364 & 44.717 \\
2001XA255 & 30.179 & 0.689427 & 165.79 & 9.373 & 50.985 \\
2003UY283 & 33.454 & 0.895188 & 193.50 & 3.506 & 63.401 \\
2003WN188 & 14.538 & 0.848719 &  55.44 & 2.199 & 26.878 \\
2004CM111 & 33.180 & 0.851053 & 191.12 & 4.942 & 61.417 \\
2004PA44  & 14.168 & 0.757876 &  53.33 & 3.430 & 24.906 \\
\enddata

\tablecomments{$a$ is semimajor axis in AU, $e$ is eccentricity, $T$ is period in years, 
     $Q$ is perihelion distance in AU, and $A$ is aphelion distance in AU.}

\end{deluxetable}

\clearpage
\begin{deluxetable}{lcccccc}
\tabletypesize{\scriptsize}
\tablecaption{Observational characteristics of asteroid candidates on 2005 April 1.\label{tbl-2}}
\tablewidth{0pt}
\tablehead{
\colhead{Asteroid} & \colhead{$R$} & \colhead{dR/dt} & \colhead{$m_{V}$} & \colhead{$H$} & 
     \colhead{$r$\tablenotemark{a}} & \colhead{$U$\tablenotemark{b}}
}
\startdata
(5335) & 20.8 & 2.09 & 26.8 & 13.3 & 12 & 2 \\
(8405) & 8.41 & 4.77 & 18.8 & 9 & 95 & 0 \\
1995SN55\tablenotemark{c} & 38.4 & -0.80 & 22.0 & 6 & 370 & n/a\tablenotemark{d} \\
1996AR20\tablenotemark{c} & 16.9 & 4.28 & 26.3 & 14 & 9 & n/a\tablenotemark{d} \\
1997MD10 & 18.1 & 7.51 & 28.8 & 16 & 4 & 1 \\
1998QJ1 & 14.0 & 5.51 & 28.3 & 16.5 & 3 & 3 \\
1998WU24 & 15.7 & 6.69 & 27.2 & 15 & 6 & 3 \\
(20461) & 13.8 & 8.36 & 25.6 & 13.8 & 9 & 0 \\
1999XS35 & 14.7 & 7.96 & 29.0 & 17.2 & 2 & 2 \\
2000HE46 & 13.2 & 8.59 & 26.4 & 14.8 & 6 & 2 \\
2001XA255 & 12.6 & -4.80 & 22.5 & 11.1 & 37 & 1 \\
2003UY283 & 6.99 & 10.3 & 24.4 & 15.3 & 6 & n/a\tablenotemark{d} \\
2003WN188 & 4.12 & 12.5 & 20.9 & 14.1 & 9 & 1 \\
2004CM111 & 6.63 & 7.50 & 22.3 & 14.2 & 9 & n/a\tablenotemark{d} \\
2004PA44 & 4.49 & 8.19 & 20.9 & 13.6 & 12 & 2 \\
\enddata
\tablenotetext{a}{Assuming an albedo of 0.05, appropriate to outer solar system objects.}
\tablenotetext{b}{See text for an explanation of the Uncertainty Parameter.}
\tablenotetext{c}{Orbit uncertainty high, object probably not observable without significant search effort.}
\tablenotetext{d}{Observations only available over a very short data arc.}

\tablecomments{$R$ is current heliocentric distance in AU, $dR/dt$ is current radial velocity
     in km/sec, $m_{V}$ current visual magnitude, $H$ is the object's absolute visual magnitude, 
     $r$ is the object's radius in km, and $U$ is the object's uncertainty parameter (see text).}

\end{deluxetable}

\end{document}